\documentstyle[psfig,epsf,conf-X,10pt]{article}
\begin{document} 
\small
\heading{
The Virtues of X--ray Clusters and the
Entropy of Cosmic Baryons}
\par\medskip\noindent
\author{%
Paolo Tozzi$^{1,2}$, Colin Norman$^{1}$
}

\address{Department of Physics and Astronomy, Johns Hopkins Univ.,
Baltimore, MD}

\address{Osservatorio Astronomico di Trieste, via Tiepolo 11, I-34131
Trieste, Italy}

\begin{abstract}
The thermodynamics of the diffuse, X-ray emitting gas in clusters and
groups of galaxies are affected by a certain amount of
non--gravitational energy input, as indicated by the scaling
properties of X--ray halos.  Such a view has been recently confirmed by
the detection of an excess entropy in the center of groups.  It is not
easy, however, to identify unambiguously the source and the epoch of
such excess entropy.  Here we show that the observations of high $z$
clusters will help in reconstructing both the dynamic and the
thermodynamic history of the diffuse cosmic baryons.
\end{abstract}

\section{An unknown factor}

The relevance of clusters of galaxies in cosmology cannot be
overstated.  The cosmological virtue of X--ray clusters mostly resides
in the fact that the X--ray properties of the Intra Cluster Medium
(ICM) can be used as a direct tracer of the total concentration of
mass.  The luminosity, and in particular the emission--weighted
temperature, offer a unique way to probe the power spectrum of
primordial density fluctuations and its evolution.

However, ten years ago it has been realized that a certain amount of
non--gravitational energy input is needed to explain the scaling
properties of X--ray halos ranging from clusters to groups
\cite{EH}\cite{K91}.  Such a non--gravitational contribution does
break the simple relation between the distribution of the ICM and that
of the total matter.  A recent detection \cite{PCN} confirmed the
existence of such extra energy, which can be conveniently quantified
in terms of an entropy excess with respect to the value expected from
gravitational processes only.  Furthermore, the specific entropy,
defined as $S\propto {\rm ln}(K)={\rm ln}(kT/\mu m_p\rho^{2/3})$,
determines the properties of both local and distant X--ray clusters.
This implies that the X--ray evolution is driven both by the dynamics
and the heating history of the gas, which, in turn, may depend on star
formation, nuclear activity, etc.  In particular, the impact of the
non--gravitational processes on the surrounding medium is unknown.
Therefore, this unpredictable factor casts a shadow on the virtue of
X--ray clusters as tracers of the distribution of matter.  From this
perspective, a physical model that includes the contribution of a
non--gravitational term will restore the reliability of X--ray
clusters, and possibly will reveal new virtues.

Here we will show that the entropy is a convenient variable to
describe the evolution of the ICM with the inclusion of a heating
contribution.  As a consequence, looking at the entropy in distant
X--ray clusters will give information on the non--gravitational
processes that affect the ICM.


\begin{figure}
\centerline{\vbox{
\psfig{figure=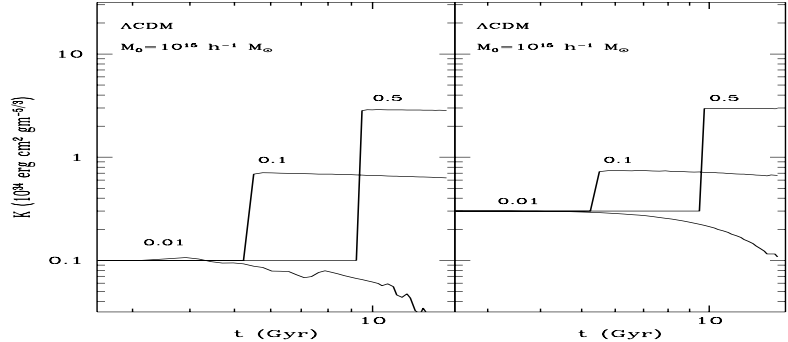,width=12cm,height=5.cm}
}}
\caption[]{\small The evolution of the entropy ($K$) of three
different baryonic shells, including respectively 1\%, 10\% and 50\%
of the total baryonic mass at $z=0$ as a function of cosmic epoch in a
$\Lambda$CDM universe.  The final mass of the halo is $10^{15} h^{-1}
M_\odot$.  The panel on the left has an initial (excess) entropy of
$K_*=0.1\times 10^{34}$ erg cm$^2$ gm$^{-5/3}$; the panel on the right
$K_*=0.3\times 10^{34}$ erg cm$^2$ gm$^{-5/3}$.}
\label{fig1}
\end{figure}

\section{The Best Record of the History of Baryons}

Let us describe the formation of X--ray halos as a spherical, smooth
accretion of shells of gas (driven by the dark matter component) with
an initial excess entropy $K_*$, which is the only free parameter.  We
can clearly distinguish three main phases in the thermodynamic
evolution of the diffuse baryons:

\begin{itemize}
\item adiabatic compression when the gas starts to be collected in the
evolving potential well and its temperature grows as $kT\propto K_*
\rho^{2/3}$;

\item shock heating as the infall velocities of the shells become
larger than the sound speed; as a consequence, the entropy of the
accreted gas shell jumps to higher values \cite{CMT};

\item further adiabatic compression of the shells enclosed within the
shock front; the shells may start to lose entropy due to the
radiative cooling, especially in the central regions.
\end{itemize}

What role is played by the initial excess entropy?  In Figure
\ref{fig1} we show the entropy history of three baryonic shells
(containing 1\%, 10\% and 50\% of the baryonic mass of a cluster of
$10^{15} h^{-1} M_\odot $ total) for two different initial values of
$K_*$.  In the first case with $K_*=0.1\times 10^{34}$ erg cm$^2$
gm$^{-5/3}$, the gas in the center of the halo becomes dense enough to
start early cooling.  Consequently, the final entropy in the center is
much lower than the initial level.  In particular, the inner shells
can cool completely and drop out from the diffuse, emitting phase.  In
the case with $K_* = 0.3 \times 10^{34}$ erg cm$^2$ gm$^{-5/3}$, the
high initial value of $K_*$ prevents most of the gas from cooling, and
a non--negligible entropy level is preserved in the center.  These
high entropy regions are responsible for the flat cores in the density
distribution, that are more extended going from clusters to groups.
This mechanism, by setting the appropriate value of $K_*$, bends the
$L$--$T$ relation from the self--similar prediction $L\propto T^2$, to
the observed average $T^3$.  Note also that the entropy level at large
radii is unaffected by the initial value, since it is dominated by
shock heating.

Summarizing, after a proper treatment of shock heating and cooling,
the entropy turns out to be the best record of the thermodynamic
history of the diffuse baryons at the scale of groups and clusters.
In particular the excess entropy and the cooling processes strongly
interfere with each other, in the sense that a non--negligible excess
entropy inhibits the radiative cooling.  Despite the simplification of
assuming a constant and homogeneous value in the external gas, this
model can reproduce many scaling properties of X--ray halos if $K_* =
(0.4\pm 0.1) \times 10^{34}$ erg cm$^2$ gm$^{-5/3}$ \cite{TN}.  Note
that the excess entropy can be generated after the collapse, but this
would require a much higher energy budget.  The last scenario is
currently under investigation.


\begin{figure}
\centerline{\vbox{
\psfig{figure=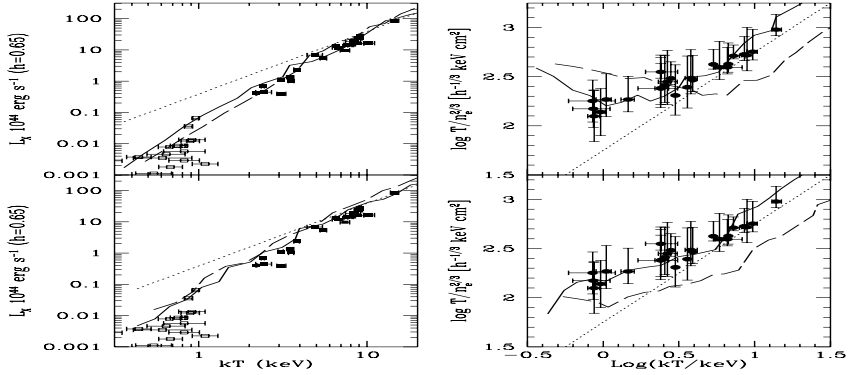,width=12.cm,height=5.cm}
}}
\caption[]{\small The $L$--$T$ relation (top left panels, data by
\cite{P96} and \cite{AE}) and the $K$--$T$ relation (top right panels,
data by \cite{PCN}) are shown for a constant $K_*=0.4\times 10^{34}$
erg cm$^2$ gm$^{-5/3}$ at $z=0$ (continuous lines) and $z=1$ (dashed
lines).  In the bottom panels, the same is shown for an evolving
external entropy  $K_*=0.8 \times 10^{34} (1+z)^{-1}$
erg cm$^2$ gm$^{-5/3}$.  Note the different definition for $K\equiv
kT/n_e^{2/3}$.}  
\label{fig2}
\end{figure}

\section{The Virtues of Clusters}

At this point, it is clear that the evolution of the $L$--$T$ relation
is affected to a large extent by the amount of the excess entropy and
its time evolution.  We recall that both the luminosity and the
emission--weighted temperature are affected, even if the $M$--$T$
relation is less dependent on the actual value of $K_*$ \cite{TN}.
However, once the non--gravitational processes can be included in the
excess entropy, the above picture unveils a new virtue of X--ray
halos.  In fact, the emission properties of clusters and groups
reflect both the dynamic and the thermodynamic history of the baryons.
Paying the price of a more complex scenario, it will be possible not
only to test the cosmology, but also, at the same time, the history of
non--gravitational processes like nuclear activity and star formation
history (e.g., coupling galaxy formation models with the evolution of
X--ray halos, see \cite{VS}\cite{WFN}\cite{MC}).

The simple case of a single value of $K_*$ in the IGM can explain many
scaling properties of local clusters, as shown in Figure
\ref{fig2} for the $L$--$T$ and the $K$--$T$ relations (where $K$ is
estimated at $r=0.1 R_{vir}$, see \cite{PCN}; note also that $L$ at
the scale of groups is computed within a radius much larger than the
$0.1 h^{-1}$ Mpc used in \cite{P96}, see discussion in \cite{TN}).
In the same Figure \ref{fig2} we show two different cases for the
evolution of the excess entropy.  We found that a constant $K_*$ will
give a roughly constant $L$--$T$, while an evolving $K_* \propto
(1+z)^{-1}$ will give similar local properties, but a higher $L$--$T$
(lower $K$--$T$) at $z\simeq 1$.  The observation of distant clusters
therefore, will unveil a large part of the history of the cosmic
baryons, and can be usefully coupled with observations in other
spectral bands in order to unambiguously identify the sources, the
time scale and the global energy budget of the non--gravitational
preheating.

{This work has been supported by NASA grant NAG 8-1133.}

\vfill

\begin{iapbib}{99}{
\bibitem{AE} Arnaud, M., \& Evrard, A.E. 1999, MNRAS, 305, 631
\bibitem{CMT} Cavaliere A., Menci N., Tozzi P. 1997, ApJ, 484, L21
\bibitem{EH} Evrard, A.~E., \& Henry, J.~P. 1991, ApJ, 383, 95
\bibitem{K91} Kaiser, N. 1991, ApJ, 383, 104 
\bibitem{MC} Menci, N., \& Cavaliere, A. 1999, ApJ, in press, astro-ph/9909259
\bibitem{P96} Ponman, T.~J., et al. 1996, MNRAS, 283, 690 
\bibitem{PCN} Ponman, T.~J., Cannon, D.~B., \& Navarro, F.J. 1999,
Nature, 397, 135 
\bibitem{TN} Tozzi, P., \& Norman, C. 1999, ApJ submitted
}
\bibitem{VS} Valageas, P., \& Silk, J. 1999, astro-ph/9907068
\bibitem{WFN} Wu, K.~K.~S., Fabian, A.~C., \& Nulsen, P.~E.~J. 1999, 
astro-ph/9907112
\end{iapbib}
\end{document}